\documentclass[12pt]{article}

\usepackage{tikz}
\usepackage{epsfig,latexsym,amsfonts,amsmath,amsthm,amssymb,amsbsy,multirow,slashed,wasysym,textcomp,subfigure,wrapfig,datetime,comment,mathtools,cancel,cite,twistor,mathrsfs}
\usepackage[hidelinks]{hyperref}
\usepackage[font={footnotesize},bf]{caption}

\def\ee{\end{equation}}
\def\be{\begin{equation}}
\def\bea{\begin{eqnarray}}
\def\eea{\end{eqnarray}}
\newcommand{\beq}{\begin{eqnarray}}
\newcommand{\eqq}{\end{eqnarray}}
 \newcommand{\badat}{\begin{alignedat}}
 \newcommand{\eadat}{\end{alignedat}}
\newcommand{\bh}{\bar{h}}
\newcommand{\eal}[1]{\be \begin{aligned} #1 \end{aligned}\end{equation}} 
\newcommand{\eqn}[1]{\be #1 \end{equation}} 
\newcommand{\eqa}[1]{\bea  #1\end{eqnarray}}
\newcommand{\CT}{\mathcal{LT}^2}
\renewcommand{\d}{\mathrm{d}}

\long\def\new#1\endnew{{\bf #1}}		
\long\def\del#1\enddel{}
\def\eps{\epsilon }
\def\del{\partial}

\usepackage{color}

\newcommand{\pink}[1]{\textcolor{\pink}{#1}}

\definecolor{dblue}{rgb}{0.2,0.50,0.80}

\def\O{\mathcal{O}}

\def\vphi{\varphi}
\def\bigma{\bar\sigma}

\def\bh{{\bar h}}
\def\bz{{\bar z}}

\def\sm{ {\bar\sigma} }
\def\s{ {\sigma} }

\newcommand{\bs}{\bar{\sigma}}

\numberwithin{equation}{section} 

\begin{document}
\begin{titlepage}
  \thispagestyle{empty}
  \begin{flushright}
  \today
    \end{flushright}
  \bigskip
  \begin{center}
	 \vskip2cm
  \baselineskip=13pt {\LARGE \scshape{Conformal Correlators\\
  \vspace{0.5em}on the Lorentzian Torus}}

	 \vskip2cm
   \centerline{Walker Melton, Atul Sharma and Andrew Strominger}
 \vskip.5cm
 \noindent{\em Center for the Fundamental Laws of Nature,}
  \vskip.1cm
\noindent{\em  Harvard University,}
{\em Cambridge, MA, USA}
\bigskip
  \vskip1cm
  \end{center}
  \begin{abstract}
The general form of a 2D conformal field theory (CFT) correlator on a Euclidean Riemann surface, Lorentzian plane or Lorentzian cylinder is well-known. This paper describes the  general form of 2- and 3-point CFT correlators on the Lorentzian torus $\CT$ which arises as  the conformal boundary of the group manifold $\SL(2,\R)$\ $\simeq \text{AdS}_3/\Z$.  We consider only generic points, thereby omitting an analysis of contact terms, which already exhibits a surprisingly rich structure. The results are relevant to celestial holography, for which the  $\CT$ at the boundary of  Klein space is  the home 
of the putative celestial CFT.
  
  \end{abstract}
%

%
\end{titlepage}
\tableofcontents

\section{Introduction}

Conformal field theory on Minkowski space $\R^{1,d-1}$ can typically be uplifted to the Einstein cylinder. The latter is the universal cover of compactified Minkowski space $(S^1\times S^{d-1})/\Z_2$. CFT correlators on the cylinder display a very rich analytic structure. The Einstein cylinder arises as a boundary of the domain of analyticity of Euclidean CFT correlators when they are analytically continued to complexified spacetime \cite{Luscher:1974ez}, as well as at the conformal boundary of AdS$_{d+1}$. Recent applications of these relations  include the study of light ray operators in Lorenztian CFT \cite{Kravchuk:2018htv}. 

An intermediate geometry between compactified Minkowski space and the Einstein cylinder is the torus $S^1\times S^{d-1}$, a double cover of the former. Here  time becomes periodic and runs along the $S^1$ factor, so it has remained unconventional to study CFT on such geometries.  An important special case is $d=2$, which we shall refer to as the Lorentzian torus $\CT$. In this paper we analyze CFT correlators on $\CT$ and find a surprisingly rich structure. 

The conformal geometry $\CT$ arises in a number of related contexts. It appears as the conformal boundary of the group manifold of $\SL(2,\R)$, which in turn is the quotient AdS$_3/\Z$ of AdS$_3$ by a unit\footnote{$i.e.$ double the AdS$_3$ light-crossing time.} global time shift. Split signature Klein space can be expressed as a foliation by AdS$_3/\Z$ hypersurfaces, whose $\CT$ boundaries are the Kleinian analogs of the Minkowski celestial sphere \cite{Atanasov:2021oyu}. Hence $\CT$ plays a central role in celestial holography, where it is known as the celestial torus. Kleinian scattering amplitudes in a basis of conformal primary states are CFT correlators on $\CT$ \cite{Pasterski:2016qvg,Pasterski:2017kqt,Pasterski:2017ylz}. 

For $\CT$, it is important to tackle the global geometry of the torus when constructing even the simplest conformal correlators like at 2 or 3 points. Even though $\CT$ is a double cover of $\R^{1,1}$, many allowed conformally covariant structures for correlators on a single Lorentzian diamond do not uplift trivially to $\CT$. They need to be regulated by careful $\im\eps$ prescriptions while preserving single-valuedness on $\CT$. Motivated by this, in this work we perform a classification of various single-valued conformal correlators at 2 and 3 points that can appear in a CFT on $\CT$. We find 2 generic allowed forms for 2-point correlators and 16 for the 3-point correlators. This rich structure contrasts with the Euclidean plane for which the 2- and 3-point correlators are uniquely determined (up to a scale) by the conformal weights. 

We focus on correlators that are single-valued for generic weights $h,\bh\in\C$.  At integer weights, new representations of the conformal group appear. These are of great interest \cite{Atanasov:2021oyu,Freidel:2022skz,Cotler:2023qwh} but not analyzed here.  We also restrict ourselves to generic points on the torus, avoiding contact terms. By `contact term' in this paper we mean generically any type of singularity (often distributional) supported only when two or more points are coincident or separated by a null geodesic.   When two or more points lie on a common light cone, distributional contact terms are allowed by conformal invariance and in general arise. Although not pursued here, a study of these terms is desirable for the application of our analysis to celestial holography, as low-multiplicity celestial amplitudes are often distributional in nature \cite{Pasterski:2017ylz}. On the other hand there is much study \cite{Crawley:2021ivb,Fan:2021isc,Sharma:2021gcz,Fan:2021pbp,Hu:2022syq,Banerjee:2022hgc,De:2022gjn,Chang:2022jut,Jorge-Diaz:2022dmy,Brown:2022miw,Casali:2022fro,Ball:2023ukj,Stieberger:2022zyk,Fan:2022vbz,Costello:2022jpg,Costello:2023hmi,Adamo:2023zeh,Melton:2022fsf,Gonzo:2022tjm,Stieberger:2023fju,Bittleston:2023bzp} of smooth celestial correlators arising from shadows or translation-breaking backgrounds which may,  in some suitable sense, be a more direct characterization of the underlying celestial CFT.   To these correlators our results apply directly. Moreover, as speculated in the concluding discussion, the distributional expressions may arise as differences of analytic ones. 

We begin in section \ref{sec:torus} by reviewing the geometry of $\CT$ from its origin as the  celestial torus in $\R^{2,2}$. Conventions are established for various local and global coordinate systems on the torus. In section \ref{sec:2pt}, we illustrate at 2 points the general class of arguments that constrain correlators on the torus. The most important constraint comes from demanding single-valuedness as one goes around various cycles of the torus. Each 2-point conformally covariant structure is defined with a choice of branch -- entering in the form of $\im\eps$ prescriptions -- and different choices pick up different monodromies as one circles the torus cycles. We assemble all the independent linear combinations of these ``building blocks'' that remain single-valued. This argument is generalized to 3 points in section \ref{sec:3pt} and higher points in section \ref{sec:higher}. In section \ref{discussion}, we end with a discussion of possible applications and future directions pertaining to celestial holography.


\section{The Lorentzian  torus}
\label{sec:torus}

This section collects our conventions. Let's start by reviewing the appearance of $\CT$ as the celestial torus at the boundary of Klein space. 

Klein space and its  conformal compactification have been studied  in \cite{Atanasov:2021oyu, Mason:2005qu}. The term \emph{Klein space} refers to flat space $\R^{2,2}$ with a metric of split, i.e., $(2,2)$ signature
\be\label{ds2}
\d s^2 = -\,(\d X^0)^2-(\d X^1)^2+(\d X^2)^2+(\d X^3)^2\,.
\ee 
In Cartesian coordinates $X^\mu$, the light cone of the origin of $\R^{2,2}$ is given by
\be
(X^0)^2+(X^1)^2 = (X^2)^2+(X^3)^2\,.
\ee
Removing the origin and quotienting this by positive rescalings $X^\mu\sim \al X^\mu$, $\al>0$, we can impose 
\be\label{s1s1}
(X^0)^2+(X^1)^2 = (X^2)^2+(X^3)^2 = 1\,.
\ee
This yields a \emph{celestial torus} $\CT=S^1\times S^1$ as the geometry of the light cone cuts. It replaces the celestial sphere when going from Lorentzian to split signature.

We can globally parametrize $\CT$ by solving \eqref{s1s1} in terms of angles $(\tau,\vphi)\in S^1\times S^1$:
\be
X^0+\im X^1 = \e^{\im\tau}\,,\qquad X^2+\im X^3 = \e^{\im\vphi}\,.
\ee
These angles have independent periodicities $\tau\sim\tau+2\pi$ and $\vphi\sim\vphi+2\pi$. But the action of the Kleinian isometry group $\SO(2,2)\sim \SL(2,\R)\times\overline{\SL}(2,\R) $ can be realized more naturally  if we work with the ``light cone coordinates''
\be
\sigma = \frac{\tau+\vphi}{2}\,,\qquad\bigma = \frac{\tau-\vphi}{2}\,.
\ee
In terms of these, the split signature metric \eqref{ds2} induces a Lorentzian conformal structure on $\CT$ represented by
\be
\d s^2_{\CT} = -\,\d\tau^2+\d\vphi^2 = -4\,\d\sigma\,\d\bigma\,.
\ee
In this conformal structure, null rays on the celestial torus are given by cycles of fixed $\sigma$ or $\bigma$.

\begin{figure}
\begin{center}

\tikzset{every picture/.style={line width=0.75pt}} 

\begin{tikzpicture}[x=0.75pt,y=0.75pt,yscale=-1,xscale=1]

\draw   (173,77.64) -- (240.4,145.04) -- (173,212.44) -- (105.6,145.04) -- cycle ;
\draw   (240.4,10.25) -- (307.79,77.64) -- (240.4,145.04) -- (173,77.64) -- cycle ;
\draw    (99,153) -- (162,217.44) ;
\draw [shift={(125.96,180.57)}, rotate = 45.65] [fill={rgb, 255:red, 0; green, 0; blue, 0 }  ][line width=0.08]  [draw opacity=0] (8.93,-4.29) -- (0,0) -- (8.93,4.29) -- cycle    ;
\draw    (315,84.56) -- (184,218.44) ;
\draw [shift={(254.05,146.85)}, rotate = 134.38] [fill={rgb, 255:red, 0; green, 0; blue, 0 }  ][line width=0.08]  [draw opacity=0] (8.93,-4.29) -- (0,0) -- (8.93,4.29) -- cycle    ;
\draw    (120,122) -- (136,122.44) ;
\draw    (261,117) -- (277,117.44) ;
\draw    (194,50) -- (210,50.44) ;
\draw    (198,45) -- (214,45.44) ;
\draw    (194,182) -- (210,182.44) ;
\draw    (198,177) -- (214,177.44) ;
\draw    (265,44) .. controls (277,29.44) and (273,52.44) .. (281,38.44) ;
\draw    (133,180) .. controls (145,165.44) and (141,188.44) .. (149,174.44) ;

\draw (255,148) node [anchor=north west][inner sep=0.75pt]    {$\sigma $};
\draw (111,189) node [anchor=north west][inner sep=0.75pt]    {$\overline{\sigma }$};

\end{tikzpicture}
\end{center}
\caption{A convenient fundamental domain for the celestial torus \label{fig:celtor}}
\end{figure}
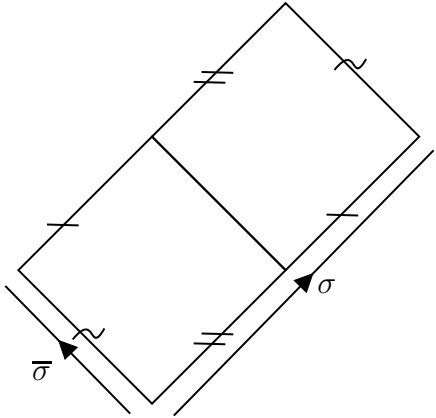

The periodicities of $\sigma$ and $\bigma$ are tied together:
\be
(\sigma,\bigma) \sim \big(\sigma+(m+n)\pi,\,\bigma+(m-n)\pi\big)\,,\qquad m,n\in\Z\,.
\ee
A convenient fundamental domain for these coordinates can be taken to be
\be \label{fdm} 
0\leq\sigma<2\pi\,,\qquad0\leq\bigma<\pi
\ee
as depicted in figure \ref{fig:celtor}. The equivalence $(\sigma,\bigma)\sim(\sigma+\pi,\bigma+\pi)$ can be used to ensure $\bigma\in[0,\pi)$. Then the equivalence $\sigma\sim\sigma+2\pi$ for fixed $\bigma$ can be used to bring $\sigma$ within $[0,2\pi)$.

The open set $\CT-\{\sigma\bigma=0\}$ is a union of two antipodally placed Lorentzian diamonds $\R^{1,1}$. On each diamond, one can introduce a pair of real and independent local coordinates
\be
z = \tan\sigma\,,\qquad\bz = \tan\bigma
\ee
that are invariant under the periodicities of $\sigma,\bigma$. Up to Weyl rescalings, the corresponding 2D metric is $-\d z\,\d\bz$. The 4D Klein group $\sim$ 2D conformal group $\SL(2,\R)\times\overline{\SL}(2,\R)$ acts by real and independent M\"obius transformations on $z$ and $\bz$. Under these transformations
\begin{equation}
    z \mapsto \frac{az+b}{cz+d}\,,\qquad \s \mapsto \arctan\left(\frac{a\tan\s+b}{c\tan\s+d}\right)
\end{equation}
where $ad-bc = 1$ \cite{Atanasov:2021oyu}. This makes $z,\bz$ ideal for studying 2D CFT on a single Lorentzian diamond.

When studying CFT on a Lorentzian torus (or the Einstein cylinder), one can again try to work with such a local description diamond by diamond. But in this description, the global structure of the torus remains obscure. For example, in celestial CFT the global structure only emerges through various indicator functions that multiply celestial amplitudes and describe the allowed scattering channels \cite{Pasterski:2017ylz}. One way to make it partially manifest is to work with boost eigenstates that diagonalize the $\Z_2$ action $X^\mu\mapsto-X^\mu$ \cite{Jorge-Diaz:2022dmy,Brown:2022miw}. In contrast, in this work we will make it completely manifest by directly studying CFT correlators on $\CT$ in global coordinates $\sigma,\bigma$. 


\section{Two-point correlators}
\label{sec:2pt}

In this section we work out the classification of conformally covariant structures for generic weight operators at 2 generic points in a Lorentzian CFT on the celestial torus. This is more constrained than  the analogous exercise on an Einstein cylinder because our time coordinate $\tau=\sigma+\bigma$ is periodic.

Let's start by analyzing 2D translation symmetry on $\CT$. Translations $\sigma\mapsto\sigma+\theta$ act as M\"obius transformations $z\mapsto(z\cos\theta+\sin\theta)/(\cos\theta-z\sin\theta)$, so we need them to be a symmetry. Let $\cO_{h,\bh}(z,\bz)$ be a conformal primary operator of weights $(h,\bh)$ in a Lorentzian CFT on $\R^{1,1}$. In global $(\sigma, \bar \sigma)$ coordinates $\cO_{h,\bh}$ uplifts to  
\be\label{Olift}
\cO_{h,\bh}(\sigma,\bigma) \vcentcolon= |\cos\sigma|^{-2h}|\cos\bigma|^{-2\bh}\cO_{h,\bh}(z,\bz)\,.
\ee
Translation invariance on $\R^{1,1}$ requires that correlators of $\cO_{h_i,\bh_i}(z_i,\bz_i)$ only depend on $z_{ij}\equiv z_i-z_j$ and $\bz_{ij}\equiv\bz_i-\bz_j$. A short calculation shows that
\be
z_{ij} = \frac{\sin\sigma_{ij}}{\cos\sigma_i\cos\sigma_j}\,,\qquad\bz_{ij} = \frac{\sin\bigma_{ij}}{\cos\bigma_i\cos\bigma_j}\,.
\ee
By conformal covariance, one finds that any correlator of the uplifted operators $\cO_{h_i,\bh_i}(\sigma_i,\bigma_i)$ will only be a function of $\sin\sigma_{ij}$, $\sin\bigma_{ij}$. These are manifestly invariant under 2D translations in $\sigma_i,\bigma_i$. The translation-breaking factors of $\cos\sigma_i$ and $\cos\bigma_i$ cancel against the Jacobians $|\cos\sigma_i|^{-2h_i}|\cos\bigma_i|^{-2\bh_i}$ coming from the uplift \eqref{Olift}.

So our task reduces to classifying conformally covariant functions of $\sin\sigma_{ij}$, $\sin\bigma_{ij}$ on the torus. We begin our analysis with 2-point functions. To be precise, we construct all the independent conformally covariant 2-point correlators on $\CT$ at generic (not null separated) points with generic (non-integral) weights.  

Our main technique is to start with conformal correlators on $\R^{1,1}$ and find combinations of them that uplift to single-valued functions on $\CT$. Consider a general 2-point correlator (for non-contact  points scale invariance requires equal weights) of an operator $\cO_{h,\bar h}(z,\bz)$  with itself, with both insertions placed in a common diamond of $\CT$. In the region $z_{12}>0$, $\bz_{12}>0$, conformal covariance dictates that  it take the form
\be
\big\la\cO_{h,\bar h}(z_1,\bz_1)\,\cO_{h,\bh}(z_2,\bz_2)\big\ra = \frac{1}{z_{12}^{2h}\bz_{12}^{2\bh}}\,,\qquad z_{12}>0\,,\;\bz_{12}>0
\ee
up to normalization. Changing to $\sigma,\bigma$ coordinates, this can be written as
\be\label{twpt}
\big\la\cO_{h,\bar h}(\sigma_1,\bigma_1)\,\cO_{h,\bh}(\sigma_2,\bigma_2)\big\ra = \frac{1}{(\sin\sigma_{12})^{2h}(\sin\bigma_{12})^{2\bh}}\,,\qquad \sin\sigma_{12}>0\,,\;\sin\bigma_{12}>0\,.
\ee
This expression has branch cuts where $\sin\sigma_{12}$ or $\sin\bigma_{12}$ vanish.  We need to provide an $\im\epsilon$ prescription if we wish to extend this 2-point function to the entire torus. 
Let $\eps>0$ be a small regulator. The independent choices of $\im\eps$ prescriptions for $(\sin\sigma_{12})^{-2h}$ and $(\sin\bigma_{12})^{-2\bh}$ are\footnote{To be completely clear about notation, the $\im\eps$ is \emph{outside} the sine functions.}
\be\label{2basis}
\frac{1}{(\sin\sigma_{12}\pm\im\eps)^{2h}}\;,\;\;\frac{1}{(\sin\bigma_{12}\pm\im\eps)^{2\bh}}\,.
\ee
In each of these, the $\im\eps$ prescription picks a branch. Note however that for any of the above choice of signs, the correlator is single valued under $2\pi$ shifts of $\sigma_k$ or $\bigma_k$, as two branch cuts with cancelling phases are crossed.\footnote{In contrast $e.g.~~ (\sin(\sigma_{12}\pm\im\eps))^{2h} $ is not single-valued for generic $h$.} 

Here, we are following the convention that expressions like $(x+\im\eps)^a$ have a branch cut along the negative imaginary axis starting at $x=-\im\eps$. Whereas $(x-\im\eps)^a$ is given a branch cut along the positive imaginary axis starting at $x=+\im\eps$. In particular, for $x\in\R-0$ and $a\in\C$, on the principal branches we obtain
\be
\lim_{\eps\to0}\,(x\pm\im\eps)^a = \begin{cases}
x^a\qquad &x>0\\
\e^{\pm\im\pi a}\,(-x)^a\qquad &x<0
\end{cases}
\ee
away from the branch point of $x^a$ at $x=0$ that shows up when $a$ is not an integer. A pair of useful identities for what follows are
\be\label{signchange}
\begin{split}
    (-x+\im\eps)^a &= \e^{\im\pi a}\,(x-\im\eps)^a\\
    (-x-\im\eps)^a &= \e^{-\im\pi a}\,(x+\im\eps)^a
\end{split}
\ee
again understood in the limit of small, positive $\eps$.

Depending on the choice of $\im\eps$ prescription used to regulate \eqref{twpt}, the resulting correlator may be single-valued on the celestial torus for a restricted set of weights or spins. We focus on  linear combinations of terms with varying $\im\eps$ prescriptions that are single-valued for generic complex weights and spins.\footnote{Non-integral spins find applications in the study of light transforms \cite{Kravchuk:2018htv,Sharma:2021gcz}.} 
For each $\im\eps$ prescription, the 2-point function is proportional to $|\sin\s_{12}|^{-2h}|\sin\bs_{12}|^{-2\bar{h}}$ with a phase depending on the signs of $\sin\s_{12}$, $\sin\bs_{12}$. 

We present the single-valued answer in terms of the conformally covariant building blocks
\be\label{2ptblocks}
G_{\eta\bar\eta}(\s_1,\sm_1;\s_2,\sm_2) = \frac{1}{(\sin\s_{12}+\im\eta\eps)^{2h}(\sin\sm_{12}+\im\bar\eta\eps)^{2\bh}}\,,\qquad\eta,\bar\eta=\pm\,.
\ee
For each $i\in\{1,2\}$, denote the combined shift by $\pi$ by the map
\be\label{linked}
\pi_i : (\s_i,\sm_i)\mapsto(\s_i+\pi,\sm_i+\pi)\,.
\ee
Our building blocks \eqref{2ptblocks} are single-valued under $2\pi$ shifts of $\s_i$ and $\sm_i$, but transform nontrivially under $\pi_1,\pi_2$. Applying the identities \eqref{signchange}, one finds
\be\label{pishift}
\pi_iG_{\eta\bar\eta} = \e^{-2\pi\im(\eta h+\bar\eta\bh)}\,G_{-\eta,-\bar\eta}\,,\qquad i=1,2
\ee
which satisfies $\pi_i^2G_{\eta\bar\eta}=G_{\eta\bar\eta}$ as expected.

We can make linear combinations of the four building blocks $G_{++},G_{+-},G_{-+},G_{--}$ and impose the constraints coming from demanding invariance under $\pi_1,\pi_2$. Note that $\pi_1$ and $\pi_2$ generate the same constraints because the expression \eqref{2ptblocks} depends only on the difference $\sigma_{12}$. Also, invariance under $\pi_1$ only generates two instead of four constraints because $\pi_1^2=1$. Using \eqref{pishift}, it follows that there are two independent combinations that remain single-valued for arbitrary $h,\bh\in\C$,
\begin{align}
  H_{++} \equiv &~~\e^{\im\pi\Delta}\,G_{++} + \e^{-\im\pi\Delta}\,G_{--}\label{2pt1}\\
  H_{+-} \equiv &~~\e^{\im\pi J}\,G_{+-} + \e^{-\im\pi J}\,G_{-+}\label{2pt2}
\end{align}
where $\Delta=h+\bh$ and $J=h-\bh$. These span a vector space of  2-point conformal correlators on $\CT$.

The branch cuts divide the domain of the correlator into two regions distinguished by the sign of the product $\sin\sigma_{12}\sin\bigma_{12}.$ (The individual signs are not invariant under $\pi_i$.) An alternate basis of two point functions can be constructed from linear combinations of \eqref{2pt1} and \eqref{2pt2} which vanish in one of the two regions. 

We note that there are other single-valued forms of the 2-point function that differ only by contact terms. A representative simple example for $h=\bar h$ is  
\be \label{rrr} 
(\sin\sigma_{12}\sin\bigma_{12}+\im\epsilon)^{-2h}\,.
\ee
This can also be generalized to incorporate spin.

The building blocks \eqref{2ptblocks} have a useful geometric interpretation. Let $\mathcal{T}^2 = S^1\times S^1$ be a torus with coordinates $\s,\sm$ running along the two $S^1$ factors, so that we impose $\s\sim\s+2\pi$ and $\sm\sim\sm+2\pi$. $\CT$ is then  a  quotient $\CT = \mathcal{T}^2/\Z_2$ where the $\Z_2$ is generated by $(\s,\sm)\sim(\s+\pi,\sm+\pi)$. So $\mathcal{T}^2$ is  a double cover of $\CT$. 
To construct single-valued objects on $\CT$, one can start with single-valued objects on $\mathcal{T}^2$. The four building blocks $G_{\eta\bar\eta}$ are precisely the single-valued possibilities for 2-point conformal correlators on $\mathcal{T}^2$, as they are invariant under $2\pi$ shifts of $\s_i$ or $\sm_i$.\footnote{The non-invariant correlators may have an interpretation -- not explored here -- as twist fields  which acquire phases around nontrivial cycles. }  To construct single-valued correlators on $\CT$, one simply computes the $\Z_2$ invariant linear combinations. This argument generalizes cleanly to higher points, as we now see for 3 points.


\section{Three-point correlators}
\label{sec:3pt}

The construction of single-valued correlators on $\CT$  proceeds similarly at 3 points. Naively uplifting the standard conformally covariant 3-point function to $\CT$ gives 
\begin{align}
    \big\langle \O_{h_1,\bh_1}(\s_1,\sm_1)\O_{h_2,\bh_2}(\s_2,\sm_2)\O_{h_3,\bh_3}(\s_3,\sm_3)\big\rangle &= \frac{1}{(\sin\s_{12})^{h_1+h_2-h_3}(\sin\s_{23})^{h_2+h_3-h_1}(\sin\s_{31})^{h_3+h_1-h_2}}\nonumber \\
    &\hspace{-3em}\times \frac{1}{(\sin\sm_{12})^{\bh_1+\bh_2-\bh_3}(\sin\sm_{23})^{\bh_2+\bh_3-\bh_1}(\sin\sm_{31})^{\bh_3+\bh_1-\bh_2}}\,.
\end{align}
As with the 2-point functions, these need to be regulated via $\im\eps$ prescriptions. The most general 3-point building blocks regulated with $\im\eps$ prescriptions are of the form
\begin{equation}
\begin{split}
    G_{\eta_{ij}, \bar\eta_{ij}} &= \frac{1}{(\sin\s_{12}+\im\eta_{12}\eps)^{h_1+h_2-h_3}(\sin\s_{23}+\im\eta_{23}\epsilon)^{h_2+h_3-h_1}(\sin\s_{31}+\im\eta_{31}\epsilon)^{h_1+h_3-h_2}} \\
    &\times \frac{1}{(\sin\sm_{12}+\im\bar\eta_{12}\eps)^{\bh_1+\bh_2-\bh_3}(\sin\sm_{23}+\im\bar\eta_{23}\eps)^{\bh_2+\bh_3-\bh_1}(\sin\sm_{31}+\im\bar\eta_{31}\eps)^{\bh_1+\bh_3-\bh_2}}\,,
    \end{split}
\end{equation}
where $\eta_{ij},\bar\eta_{ij}=\pm$. These are invariant under the shifts $(\s_i,\sm_i) \mapsto (\s_i + 2\pi m,\sm_i+2\pi n)$ for integral $m,n$, so are single-valued functions on the double cover $\mathcal{T}^2$ of $\CT$. 

There is an algorithmic way to generate all possible combinations of these building blocks that descend to $\CT$. Start with any block $G_{\eta_{ij},\bar\eta_{ij}}$. By summing the images of $G_{\eta_{ij},\bar\eta_{ij}}$ under each of the linked periodicities \eqref{linked}, one can construct the combination
\begin{equation}\label{H3}
   H_{\eta_{ij},\bar\eta_{ij}}=  G_{\eta_{ij},\bar\eta_{ij}} + \pi_1 G_{\eta_{ij},\bar\eta_{ij}} + \pi_2 G_{\eta_{ij},\bar\eta_{ij}} + \pi_3G_{\eta_{ij},\bar\eta_{ij}}\,.
\end{equation}
This is single-valued on the celestial torus because $\pi_i^2 = 1$ and the triple shift $\pi_1\pi_2\pi_3$ acts trivially on $G_{\eta_{ij},\bar{\eta}_{ij}}$ due to overall translation invariance on the torus. 
Since \eqref{H3} constructs a single-valued function out of every $G_{\eta_{ij},\bar\eta_{ij}}$, it certainly provides a basis for single-valued 3-point functions on $\CT$, if only an overcomplete one.

Each combination of the form \eqref{H3} involves four distinct $G$'s. So the space of independent combinations is $2^6/4=2^4$ dimensional. A minimal basis is provided by the 16 functions
\begin{equation}\label{stn}
    H_{\eta_{ij},+++}\,,\; H_{\eta_{ij},---}
\end{equation}
which define independent 3-point functions on $\CT$. Any 3-point function on $\CT$ can be written as a sum of these 3-point functions up to possible contact terms. Other choices of bases could of course be relevant for specific applications.

The domain of the 3-point function is divided by branch cuts into 16 regions characterized by the 4 signs of the invariant products 
\be \sin{\sigma_{12}}\sin{\bigma_{12}}, ~~~\sin{\sigma_{23}}\sin{\bigma_{23}}, ~~~\sin{\sigma_{31}}\sin{\bigma_{31}}, ~~~\sin{\sigma_{12}}\sin{\sigma_{23}}\sin{\sigma_{31}}.\ee
16 linear combinations of the 16 basis elements \eqref{stn} can be found which are nonvanishing only in any one of the 16 regions. These linear combinations provide an alternate basis for three point functions on $\CT$. 


\section{Higher-point correlators}
\label{sec:higher}

In this section, we count the number of independent choices of $\im\epsilon$ prescriptions for an $n$-point function. Noting that the conformal cross ratios
\begin{equation}
    \frac{\sin\s_{ij}\sin\s_{kl}}{\sin\s_{ik}\sin\s_{jl}}\;,\;\; \frac{\sin\sm_{ij}\sin\sm_{kl}}{\sin\sm_{ik}\sin\sm_{jl}}
\end{equation}
are themselves single-valued on the celestial torus, we need to specify a choice of $\im\epsilon$ prescription for the conformally covariant prefactor of the $n$-point function. 

Deforming 
\begin{equation}
    \begin{split}
        &\sin\s_{ij} \mapsto \sin\s_{ij} + \im\eta_{ij}\epsilon\\
        &\sin\sm_{ij}\mapsto\sin\sm_{ij}+\im\bar\eta_{ij}\epsilon
    \end{split}
\end{equation}
gives a total of $2^{2\binom{n}{2}}$ choices of $\im\epsilon$ prescriptions leading to a single-valued function on the double cover of the torus. To assemble these into single-valued functions on the celestial torus, we can sum over the independent $\pi$ shifts. Because the overall function is invariant under global translations, the shift $\pi_{i}$ is equal to the shift $\prod_{j\ne i}\pi_j$, so that we need to sum $2^{n-1}$ terms to construct a single-valued function. Therefore, for an $n$-point function, there are 
\begin{equation}
    N_n = 2^{2\binom{n}{2}-n+1}
\end{equation}
independent choices of $\im\epsilon$ assignments that generate a single-valued function of $\CT$. That is, we find $N_n$ distinct types of non-contact single-valued $n$-point conformal correlators on the celestial torus.


\section{Discussion}
\label{discussion}

This note is largely concerned with the problem of building single-valued conformal correlators on $\CT$. We classified, for generic weights up to  contact terms, correlators that obey the standard constraints of conformal symmetry on each Lorentzian diamond and uplift to single-valued functions on $\CT$.  Generic correlators at separated $\s_i,\sm_i$ can be written as linear combinations of these correlators.
We conclude with speculations on the smooth 3-point correlators \eqref{stn} as the fundamental building blocks for celestial CFT, and their relation to distributional 3-point correlators \cite{Pasterski:2017ylz} --  not falling under this classification -- arising  from  direct  Mellin transforms.

 The distributional singularities encountered in \cite{Pasterski:2017ylz} are relics of momentum-conserving delta functions. They are consistent with conformal invariance  and implied by translation invariance \cite{Law:2019glh}. In recent  literature, various approaches have been employed for avoiding distributional celestial amplitudes while staying in flat space, either using  integral transforms like shadow and light transforms or expanding around a translation non-invariant background \cite{Crawley:2021ivb,Fan:2021isc,Sharma:2021gcz,Fan:2021pbp,Hu:2022syq,Banerjee:2022hgc,De:2022gjn,Chang:2022jut,Jorge-Diaz:2022dmy,Brown:2022miw,Casali:2022fro,Ball:2023ukj,Stieberger:2022zyk,Fan:2022vbz,Costello:2022jpg,Costello:2023hmi,Adamo:2023zeh,Melton:2022fsf,Gonzo:2022tjm,Stieberger:2023fju,Bittleston:2023bzp}. The idea is that the underlying celestial CFT may be more easily understood in these smooth contexts.

While we studied smooth correlators at generic points in this paper, we nevertheless envisage that our results may be useful for the study of contact terms. 
The basic idea comes from the observation that 
\be\label{dfg}
-2\pi\im\,\delta(x) = \frac{1}{x+\im\eps} - \frac{1}{x-\im\eps}\,.
\ee
This identity and generalizations thereof suggest that the contact form of the amplitudes may arise from taking the difference of the more familiar looking, everywhere nonvanishing, amplitudes such as the individual basis elements in \eqref{stn}. Indeed any translation invariant combination of 3-point amplitudes $must$ be a contact term \cite{Law:2019glh}.\footnote{Note however that no linear combination of the basis elements in \eqref{stn} vanishes in all 16 regions. An additional term with a modified $\im\epsilon$ prescription such as the one in \eqref{rrr} would be needed.}

How would such differences between amplitudes naturally arise? One way is via the construction of celestial amplitudes as AdS-Witten diagrams \cite{Casali:2022fro, Iacobacci:2022yjo}  associated to the distinct hyberbolic foliation regions spacelike and timelike separated from the origin. Each such region gives a smooth conformally, but non-translationally, invariant `semi-amplitude' which must be of the form \eqref{stn} at generic points. On the other hand (for 3 points) the translation invariant sum of semi-amplitudes must be  a contact term. Related work appears in 
\cite{Melton:2023} which constructs Lorentz but not translation invariant Rindler-like  `hyperbolic vacua'. Celestial amplitudes in some of these vacua are expected to be smooth.
On the other hand amplitudes in the Poincar\'e vacuum can be reexpressed as sums of these. Compatibility of both these observations may be possible through higher-point generalizations of the identity \eqref{dfg}. We hope to pursue this in the future. 

\section*{Acknowledgements}

This work was supported by DOE grant de-sc/0007870, NSF GRFP grant DGE1745303, the Simons Collaboration on Celestial Holography and the  Gordon and Betty Moore Foundation and the John Templeton Foundation via the Black Hole Initiative.

\bibliographystyle{JHEP}
\bibliography{cpb}

\end{document}